\documentclass[a4paper,twocolumn,preprintnumbers,superscriptaddress,accepted=2024-02-02]{quantumarticle}
\pdfoutput=1
\usepackage{amsmath}
\usepackage{amssymb}
\usepackage{graphicx}
\usepackage{hyperref}
\usepackage{physics}
\usepackage{xcolor}
\usepackage{bm}
\usepackage{xspace,slashed}
\usepackage{multirow}
\usepackage{enumitem}
\usepackage{fontawesome}
\usepackage{utfsym}
\usepackage[OT1]{fontenc}
\usepackage[numbers,sort&compress]{natbib}

\usepackage{array}

\begin{document}

\title{Qibolab: an open-source hybrid quantum operating system}


\newcommand{\MIaff}{TIF Lab, Dipartimento di Fisica, Universit\`a degli Studi di
  Milano, Italy}

\newcommand{\INFNUNIMI}{INFN, Sezione di Milano, I-20133 Milan, Italy.}

\newcommand{\UNIMIB}{Dipartimento di Fisica, Universit\`a di Milano-Bicocca, I-20126 Milano, Italy.}

\newcommand{\INFNUNIMIB}{INFN - Sezione di Milano Bicocca, I-20126 Milano, Italy.}

\newcommand{\TII}{Quantum Research Center, Technology Innovation Institute, Abu Dhabi, UAE.}

\newcommand{\CERNaff}{CERN, Theoretical Physics Department, CH-1211
  Geneva 23, Switzerland.}

\newcommand{\UB}{Departament de F\'isica Qu\`antica i Astrof\'isica and Institut de Ci\`encies del Cosmos (ICCUB), Universitat de Barcelona, Barcelona, Spain.}

\newcommand{\Rome}{Istituto Nazionale di Fisica Nucleare (INFN), Sezione di Roma, Rome, Italy}

\newcommand{\SAP}{La Sapienza University of Rome, dep. of Physics, Rome, Italy}

\newcommand{\CQT}{Centre for Quantum Technologies, National University of Singapore, Singapore.}

\newcommand{\NTU}{Division of Physics and Applied Physics, School of Physical and Mathematical Sciences, Nanyang Technological University, 21 Nanyang Link, Singapore 637371, Singapore.}

\newcommand{\Qibolab}{\texttt{Qibolab}\xspace}
\newcommand{\Qibo}{\texttt{Qibo}\xspace}
\newcommand{\Qibocal}{\texttt{Qibocal}\xspace}
\newcommand{\Qibosoq}{\texttt{Qibosoq}\xspace}

\newcommand{\cell}[2][c]{\begin{tabular}[#1]{@{}c@{}}#2\end{tabular}}

\author{Stavros Efthymiou}
\affiliation{\TII}

\author{Alvaro Orgaz-Fuertes}
\affiliation{\TII}

\author{Rodolfo Carobene}
\affiliation{\UNIMIB}
\affiliation{\INFNUNIMIB}
\affiliation{\TII}

\author{Juan Cereijo}
\affiliation{\TII}
\affiliation{\UB}

\author{Andrea Pasquale}
\affiliation{\TII}
\affiliation{\MIaff}
\affiliation{\INFNUNIMI}

\author{Sergi Ramos-Calderer}
\affiliation{\TII}
\affiliation{\UB}

\author{Simone Bordoni}
\affiliation{\TII}
\affiliation{\Rome}
\affiliation{\SAP}

\author{David Fuentes-Ruiz}
\affiliation{\TII}

\author{Alessandro Candido}
\affiliation{\MIaff}
\affiliation{\INFNUNIMI}
\affiliation{\CERNaff}

\author{Edoardo Pedicillo}
\affiliation{\TII}
\affiliation{\MIaff}
\affiliation{\INFNUNIMI}

\author{Matteo Robbiati}
\affiliation{\MIaff}
\affiliation{\CERNaff}

\author{Yuanzheng Paul Tan}
\affiliation{\NTU}

\author{Jadwiga Wilkens}
\affiliation{\TII}

\author{Ingo Roth}
\affiliation{\TII}

\author{Jos\'e Ignacio Latorre}
\affiliation{\TII}
\affiliation{\CQT}
\affiliation{\UB}

\author{Stefano Carrazza}
\affiliation{\CERNaff}
\affiliation{\MIaff}
\affiliation{\INFNUNIMI}
\affiliation{\TII}

\begin{abstract}
  We present \Qibolab, an open-source software library for quantum hardware
  control integrated with the \Qibo quantum computing middleware framework.
  \Qibolab provides the software layer required to automatically execute
  circuit-based algorithms on custom self-hosted quantum hardware platforms.
  We introduce a set of objects designed to provide programmatic access to
  quantum control through pulses-oriented drivers for instruments, transpilers
  and optimization algorithms. \Qibolab enables experimentalists and developers
  to delegate all complex aspects of hardware implementation to the library so
  they can standardize the deployment of quantum computing algorithms in a
  extensible hardware-agnostic way, using superconducting qubits as the first
  officially supported quantum technology.
  We first describe the status of all components of the library, then we show
  examples of control setup for superconducting qubits platforms.
  Finally, we present successful application results related to circuit-based algorithms.
\end{abstract}

\maketitle

\tableofcontents

\section{Introduction}

A successful deployment of quantum computing algorithms requires quantum
hardware and middleware software dedicated to instrument control for specific
quantum platform technologies.

The goal of middleware is to provide standardized software tools which abstract
heterogeneous software interfaces from high-level applications. From quantum
computing algorithms based on the quantum circuit paradigm, to low-level driver
instructions dedicated to a specific experimental setup including instruments.
A proper implementation of middleware software accelerates research from theory
to experiments by reducing the amount of effort and expertise required to
operate a quantum platform and develop novel quantum algorithms.

Nowadays, the major challenges of middleware, as a research accelerator, include
the need of standard code procedures for quantum control algorithms, calibration
and characterization, all extensively tested and reviewed. This software should
be designed in such a way that it could be reused by similar experiments in
multiple research laboratories dedicated to quantum hardware design and fabrication.
Therefore, one of the expected positive side effects of the development of
middleware is the generation of a database of algorithms and procedures built
and maintained by a large research community. As an example, it is possible to
find similar cases in other research fields such as data analysis
tools~\cite{BRUN199781} and Monte-Carlo event generators~\cite{Alwall_2014} for
high-energy physics and artificial intelligence~\cite{tensorflow2015-whitepaper}.

Since the beginning of 2020, despite the growing interest in quantum computing
and the recent developments in quantum hardware platforms, we have observed the
lack of a standard middleware open-source framework dedicated to self-hosted
quantum platforms. There are software libraries dedicated to quantum
computing such as \texttt{Cirq}~\cite{cirq} and \texttt{TensorFlow
Quantum}~\cite{tfq} from Google, \texttt{Qiskit}~\cite{Qiskit} from IBM,
\texttt{PyQuil} from Rigetti~\cite{pyquil}, among
others~\cite{intelqs,qcgpu,qulacs,Jones_2019,10.1007/978-3-319-27119-4_17,Steiger_2018,qsharp_2017,zulehner2017advanced,pednault2017paretoefficient,PhysRevLett.116.250501,DERAEDT2007121,Fried_2018,Villalonga_2019,luo2019yaojl,bergholm2018pennylane,10.1145/3310273.3323053,10.1007/978-3-030-50433-5_35,Jones_2020,Chen_2018,EasyChair:4050,meyerov2020simulating,moueddene2020realistic,wang2020quantum}.
However, many of these software libraries have been promoted just to grant
users access to freeware and/or commercial cloud-based platforms, hence no
full-stack open-source library for quantum algorithms, from simulation to
quantum hardware control was available.
Moreover, specialized quantum hardware solutions such as \texttt{QCodes}~\cite{qcodes},
\texttt{PyCQed}~\cite{pycqed} or \texttt{Labber}~\cite{labber} offer too rigid a
structure to seamlessly incorporate all the other essential features that a full-stack solution requires.
Therefore, we started developing \Qibo~\cite{Efthymiou_2021,Efthymiou_2022,Carrazza_2023,stavros_efthymiou_2023_7736837},
an open-source middleware framework for quantum computing, by establishing an
international collaboration network involving laboratories in universities and
research institutions located in Europe, Asia and America.

In this manuscript we present for the first time
\Qibolab~\cite{stavros_efthymiou_2023_7748527}, a software library which unlocks
\Qibo's potential to execute quantum algorithms on self-hosted quantum hardware
platforms. We provide a dedicated application programming interface (API) for
quantum circuit design, qubit calibration, instrument control through arbitrary
pulses, driver operations including sweepers and transpilation into a given
platform topology using its native gates. A successful implementation of \Qibo
will deliver to the research community a first prototype of extensible quantum
hardware-agnostic open-source \textit{hybrid quantum operating system}, fully
tested and benchmarked on superconducting platforms.

The paper is organized as follows. In Sec.~\ref{sec:qibo} we describe the
project status, design and modules. Then, in Sec.~\ref{sec:drivers} we present a
detailed overview of the \Qibolab library for version \texttt{0.1.0}. In
Sec.~\ref{sec:application} we show examples of applications involving
superconducting qubit platforms. Finally, in Sec.~\ref{sec:outlook} we draw our
conclusion and discuss about future development directions.

\section{Project overview and specification}
\label{sec:qibo}

In this section we summarize the status of \Qibo in the release \texttt{0.2.0} by
describing the software design, the latest features implemented in modules and
tools, including simulation, hardware control and calibration.
The aim of this section is to provide an updated high-level description overview
of the project, following up the previous releases documented in
Refs.~\cite{Efthymiou_2021} and~\cite{Efthymiou_2022}.

For an in depth technical description of the \Qibolab library and its software features
we invite the reader to proceed to Sec.~\ref{sec:drivers}
and~\ref{sec:application}.

\subsection{Software design}

\begin{figure*}
  \center
  \includegraphics[width=0.9\textwidth]{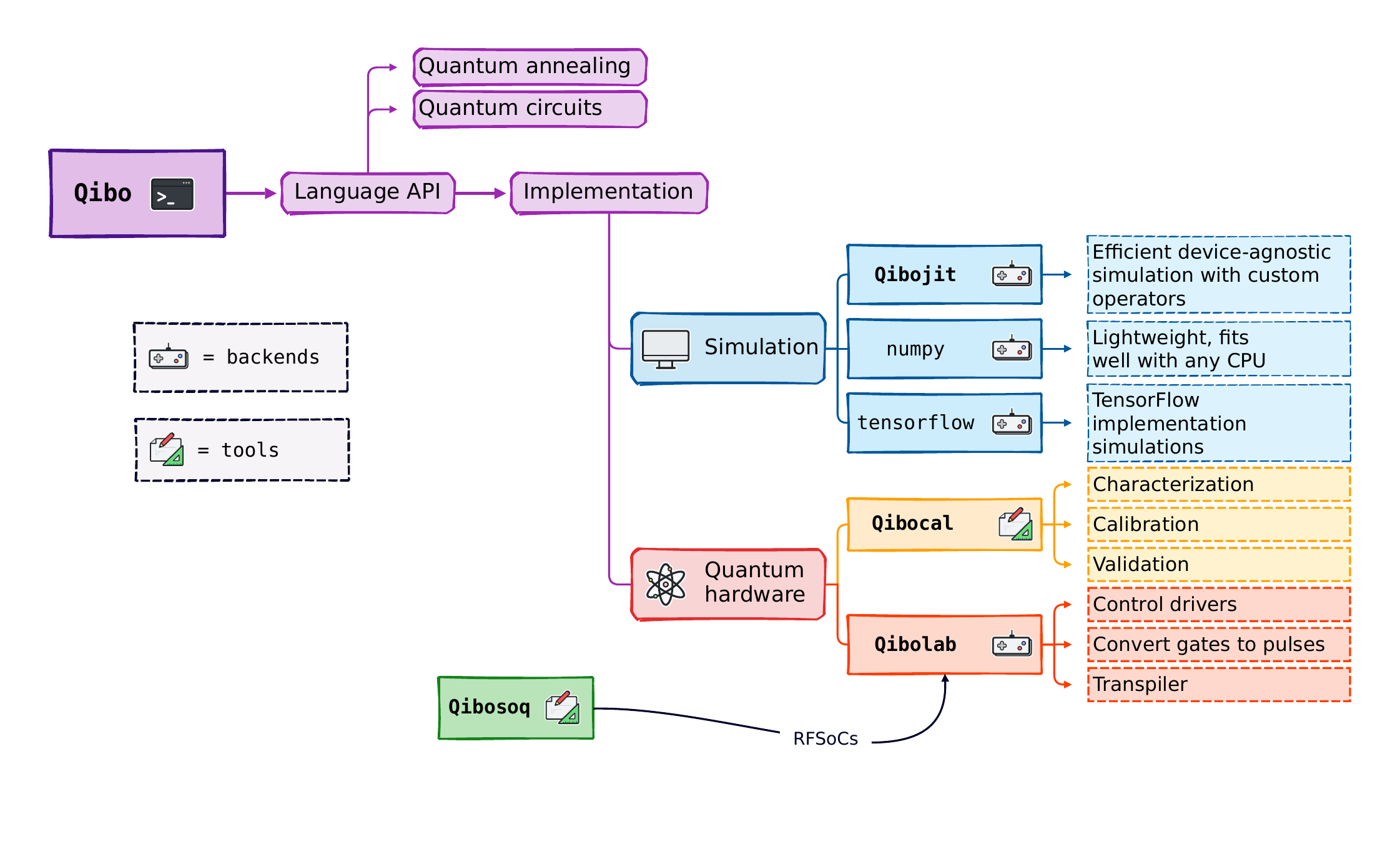}
  \caption{\label{fig:design}Schematic overview of \Qibo software components,
  including backends and tools, for release \texttt{0.2.0}.}
\end{figure*}

In Fig.~\ref{fig:design} we schematically show \Qibo's layout. The framework
is divided into two blocks: the language API and the backends implementation for
execution on various classical or quantum hardware.

The API contains a set of high-level interfaces for fast
prototyping of quantum computing algorithms based on circuit and adiabatic
paradigms adopting Python as programming language.

The quantum circuit API implements primitives for exact quantum state manipulation,
circuit model initialization with single and two-qubit
gates, as well as more complex operations such as Toffoli gates and gate fusion.
This API also includes an exhaustive interface to perform final state measurements through shots.
Furthermore, dedicated functions are available for noisy quantum simulation on
classical hardware. The user has the possibility to build custom noise models
through channels such as Kraus channel operators~\cite{preskill_channels},
a multi-qubit noise channel that applies Pauli operators with given
probabilities, an $n$-qubit depolarizing quantum error channel,
single-qubit thermal relaxation error channels or readout and single-qubit
reset channels.
Error mitigation techniques for quantum circuits are also available with the
following algorithms: Zero Noise Extrapolation (ZNE)~\cite{PhysRevA.102.012426},
Clifford data regression (CDR)~\cite{Sopena_2021}, randomized
readout~\cite{van_den_Berg_2022} and Variable Noise CDR
(vnCDR)~\cite{Sopena_2021}.

These circuit-based primitives are complemented by a database of
circuit-based models such as the quantum fourier transform (QFT)~\cite{QFT},
variational quantum eigensolver (VQE)~\cite{vqe}, adiabatically assisted
variational quantum eigensolver (AAVQE)~\cite{aavqe}, quantum approximate
optimization algorithm (QAOA)~\cite{qaoa}, feedback-based algorithm for quantum
optimization (FALQON)~\cite{Magann_2022}, style-based quantum generative
adversarial networks (style-QGAN)~\cite{Bravo_Prieto_2022}, Grover's
algorithm~\cite{grover1996fast} and the travelling salesman problem
(TSP)~\cite{Hadfield_2019}.
In addition to these features, \Qibo also provides a set of optimizers and
callbacks for variational circuit optimization and a module with quantum information
primitives.

Our annealing module API provides algorithms for time evolution of quantum states,
symbolic and numeric matrix-based Hamiltonian allocation and adiabatic
evolution~\cite{farhi2000quantum}.
In order to accelerate the initialization of Hamiltonians, \Qibo provides a
database of pre-coded models including the Heisenberg \textit{XXZ}, the
non-interacting Pauli-\textit{X/Y/Z}, the transverse field Ising model (TFIM) and
the max cut Hamiltonian.

In~\cite{Efthymiou_2021}, practical examples illustrating the implementation of
the aforementioned features of \Qibo are provided. Additional examples are provided in the \Qibo documentation~\cite{qibo-examples}.

From the implementation point of view \Qibo provides multiple \emph{execution backends}
which are responsible for the conversion and execution of the primitives on
different hardware. Each backend inherits from an abstract interface which
determines the set of methods that must be implemented in order to execute the
primitives of the language API.

At the current stage, we support simulation
backends on classical hardware and, through \Qibolab, the same high-level code
can be executed directly on quantum hardware. In practical terms, we can consider
\Qibolab as an actual hardware backend, once a specific platform is selected by the user.
Furthermore, this modularity opens the possibility to create further tools which
rely on \Qibo and its backends. For example, tools for quantum chemistry,
multi-qubit calibration routines, benchmarking, machine learning
inspired algorithms and others,
as well as the addition of further backends for simulation or hardware execution.

\subsection{Classical quantum simulation}

Simulation is a crucial part of quantum computing research,
particularly in the current Noisy Intermediate-Scale Quantum (NISQ)~\cite{Preskill_2018}
era, where exact results from simulation can be used for validating algorithms
or implementing error mitigation routines.

In \Qibo, both gate-based and adiabatic quantum computation paradigms can be
simulated on classical hardware. Thanks to its modularity, quantum algorithms
can be deployed on three different simulation backends, which are designed to
meet specific needs, as represented in Fig.~\ref{fig:design}.
In this section we summarize the advantages and limitations of each backend currently
available in \Qibo: \texttt{numpy}, \texttt{tensorflow} and \texttt{qibojit}.
We also highlight which backend is best suited depending on the application.

\begin{figure*}
  \center
  \includegraphics[width=0.77\textwidth]{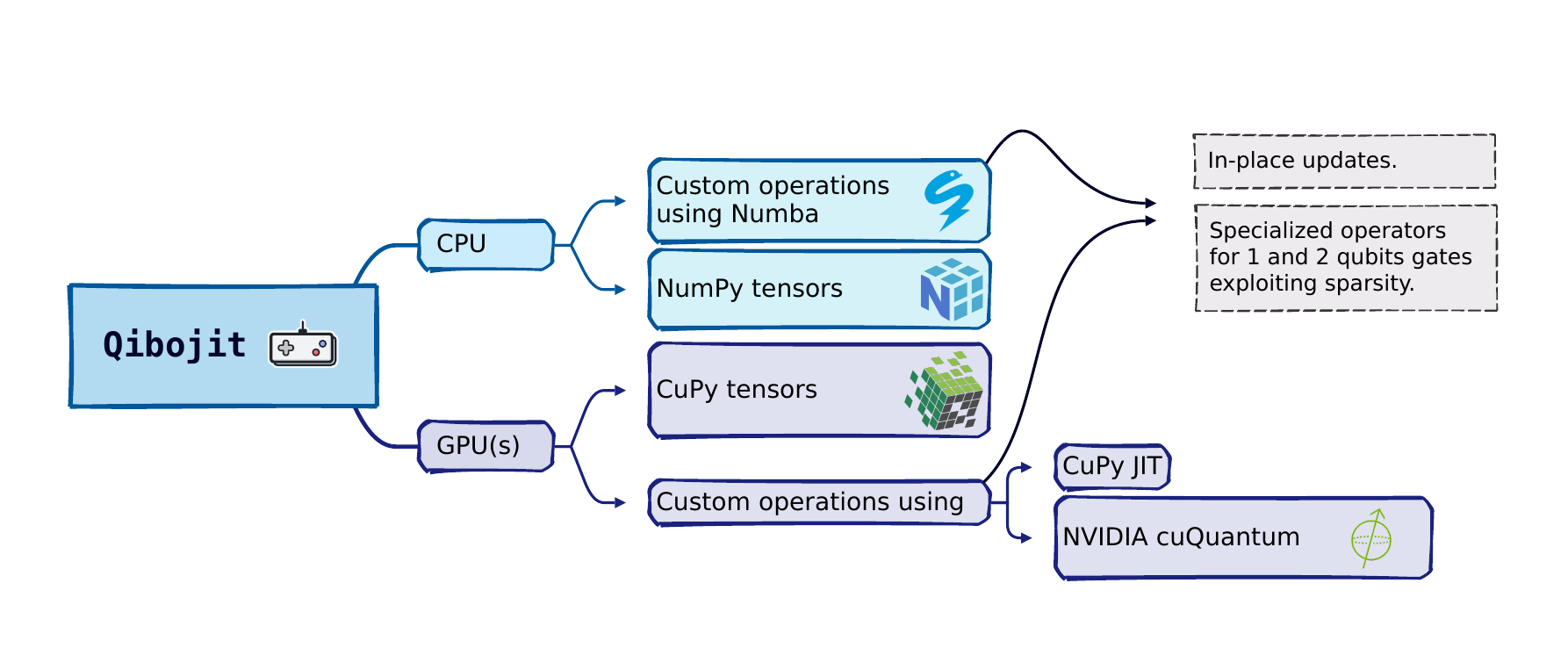}
  \caption{Schematic description of the \texttt{qibojit} backend features.}
  \label{fig:qibojit}
\end{figure*}

The \texttt{numpy} backend is based on NumPy's primitives~\cite{numpybook}, as
explained in more detail in~\cite{Carrazza_2023}. It is a lightweight backend,
which supports single-threaded CPU simulations with a moderate performance. This
setup is usually recommended for circuits up to 20 qubits. The importance of
this backend lies in its broad compatibility with many classical system
architectures, including \texttt{arm64}, which makes it a safe and stable
choice, especially in development contexts, e.g.~in laboratories where quantum
platforms are being installed and tested.

The second backend is based on TensorFlow~\cite{tensorflow2015-whitepaper}
primitives. Similarly to the \texttt{numpy} backend, it can be used for
tackling problems involving a limited number of qubits, although it allows to
perform quantum simulation on multi-threading CPU and single-GPU.
The \texttt{tensorflow} backend inherits TensorFlow's optimization
routines, including state-of-the-art gradient-based optimizers. This feature is
particularly useful in the context of Quantum Machine Learning (QML), where
automatic differentiation routines can be exploited for training hybrid
quantum-classical machine learning models~\cite{Bravo_Prieto_2022}.

Within the optimization module of \Qibo, we have implemented a function that
uses the automatic differentiation provided by TensorFlow to execute gradient-based
optimization strategies, namely \texttt{qibo.optimizers.sgd}. This function can be
customized according to the developer needs on top the features offered
by TensorFlow itself. To execute this function the usage of the \texttt{tensorflow}
backend is required.

Typically, in Machine Learning, gradients are calculated through the
Back-Propagation (BP) algorithm~\cite{Rumelhart1986LearningRB}, which requires
saving copies of matrices and vectors during the process. Since TensorFlow uses
this method, the \texttt{tensorflow} backend requires copies of the state vector
during simulation, which increases memory consumption and reduces performance.

The third backend is \texttt{qibojit}~\cite{Efthymiou_2022}, a high-performance
simulation backend which combines Just-In-Time (JIT) compilation with the
definition of custom operators for state vector manipulation. Here, the action
of quantum gates is optimized, by considering matrix properties like sparsity
and symmetries, and by avoiding allocating new copies of matrices and vectors,
which are instead modified in-place. The \texttt{qibojit} structure is shown in
Fig.~\ref{fig:qibojit}, which shows the specific implementation adopted for CPU
and GPU(s) environments.

Multi-threading CPU, GPU and multi-GPU configurations are supported by \texttt{qibojit}.
Simulation on CPUs are based on NumPy tensors and accelerated with
Numba~\cite{lam2015numba}, while for GPU and multi-GPUs executions, we adopt
CuPy~\cite{cupy_learningsys2017}. For GPUs two different acceleration strategies
are implemented. First, we exploit the Cupy's \texttt{RawKernel} method, thanks
to which we can write custom CUDA kernels in C++ and seamlessly import them in
Python. The second accelerated simulator is implemented using primitives from
NVIDIA cuQuantum~\cite{the_cuquantum_development_team_2023_7806810}. The
\texttt{qibojit} backend is the suggested choice for simulating systems with a
large number of qubits.

In Section 3.1 of~\cite{Efthymiou_2022}, we have conducted benchmarking tests using
\Qibo on different classical hardware, focusing on significant quantum circuits like
Quantum Fourier Transform~\cite{coppersmith2002approximate} and
Bernstein-Vazirani~\cite{bv_circ}, as the number of qubits increases.
These benchmarks also include comparisons of \Qibo's performance with other public
quantum computing libraries.

Recognizing the importance of simulation, even as technology advances and the
quality of quantum devices improves, we plan to improve \Qibo from a
simulation perspective. With this in mind, we are working on the development of
new backends, supporting multi-node distribution of state vector simulation
and, by changing simulation method completely, the construction of a tensor networks~\cite{biamonte2017tensor,
Yuan_2021, Huggins_2019, ORUS2014117, biamonte2020lectures} backend.

\subsection{Quantum hardware support}
In the previous section we have shown how \Qibo can be used for quantum circuit simulation.
Although simulation is a useful tool for testing and profiling quantum algorithms,
we are still mainly interested on deploying such algorithms on quantum processors to show
the advantages of this technology \cite{48651}.

Quantum computers can be implemented using several quantum systems, including superconducting
circuits \cite{PRXQuantum.2.040202}, trapped ions \cite{RevModPhys.75.281} or neutral atoms
\cite{Henriet_2020} among others.
In this paper, we focus on superconducting devices, but \Qibolab provides an extensible abstraction library
to accommodate other quantum technologies, the only precondition is that the experimental
setup should be composed by instruments that communicate with each other and the QPUs.
As described broadly in Sect.~\ref{sec:abstractions}, it is possible to mirror any experimental
configuration by inheriting the \texttt{Platform} class, and deploying the suitable \texttt{Instruments}
and \texttt{Qubits} methods, specifying all connections through \texttt{Channels}
class.
\emph{Transmons}~\cite{Koch_2007} are one possible implementation of qubits through
superconducting devices, which are weakly anharmonic oscillators made using Josephson junctions
~\cite{Josephson:1962zz}.
To perform measurements, transmons are dispersively coupled to superconducting resonators
which are in turn coupled to a microwave transmission line.

Gates are implemented by coupling qubits through microwave drive and flux \cite{Koch_2007}
lines that carry control pulses with precise amplitude and duration.

As shown in Fig.~\ref{fig:design}, \Qibolab includes all the necessary components
to construct a backend for the deployment of quantum algorithms on self-hosted quantum
processing units (QPUs).
The addition of such backend is facilitated by \Qibo's modular layout~\cite{Carrazza_2023}
which enables users to create custom backends with minimum effort. For the particular case
of a \emph{hardware} backend this feature allows us to focus only on low-level components.

\Qibolab provides an API to define \texttt{Pulse} objects, able to perform
low-level manipulations such as executing a specific sequence of microwave pulses
similarly to other libraries~\cite{Alexander_2020,Silv_rio_2022,LabOneQ,qm_qua}.
Through this interface it is possible to code easily both experiments
and calibration protocols. Such abstraction is quite practical given that instruments
may have different definitions for specific waveforms.

Another key element listed in Fig.~\ref{fig:design} is the presence of drivers to control
and interface \Qibo with different instruments. To generate the appropriate microwave pulses
needed to perform quantum gates, a common approach is to use Arbitrary Waveform Generators
(AWG), digital to analogue (DAC) and analogue to digital converters (ADC) which are nowadays
available through Field Programmable Gate Arrays (FPGAs).
All these devices usually provide libraries or packages to control them, e.g.,
Qblox Instruments~\cite{QbloxInstruments}, Qcodes~\cite{qcodes} and LabOneQ~\cite{LabOneQ}.
Despite such heterogeneity, \Qibolab defines a common
interface to properly expose package methods required to control QPUs.

\begin{figure}
  \center
  \includegraphics[width=0.75\columnwidth]{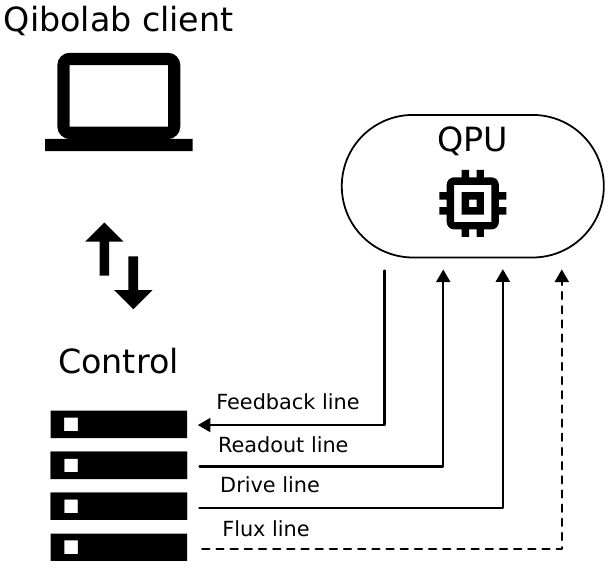}
  \caption{\label{fig:qubit_scheme}Basic setup of a self-hosted QPU. The host computer running \Qibolab communicates with the different electronics used to control a QPU.}
\end{figure}

Finally \Qibolab takes care of all the necessary operations to prepare
the execution of quantum circuits on a fully characterized device. Among these,
there is a transpilation step of circuits to the native gates supported by the
quantum processor and a compilation step to convert these gates to pulses.
Sect.~\ref{sec:transpiler} presents a more precise description of the transpilation step.

Fig.~\ref{fig:qubit_scheme} shows a basic laboratory setup for controlling a
QPU. \Qibolab is running on a host computer, which communicates, typically via a
network protocol, with the control electronics used for pulse generation. These
electronics are connected to the QPU via different channels: the readout and
feedback channels in a closed loop for measuring the qubit, the drive channel
for applying gates and, for flux-tunable qubits, the flux channels for tuning
their frequency.

For a more detailed description of the \texttt{Qibolab} backend,
we invite the reader to check Sect.~\ref{sec:abstractions}.

\subsection{Hardware characterization}

While the API provided by \Qibolab enables full control over the electronics
interacting with the qubits, this alone is not sufficient for operating a
quantum computer. This is because the accurate fine-tuning and calibration of
control waveform parameters are crucial requirements for quantum hardware to
work successfully~\cite{naghiloo2019introduction}.

Within the \Qibo environment, \Qibocal ~\cite{andrea_pasquale_2023_7662185,
pasquale2023opensource} offers the necessary tools for calibrating,
characterizing, and validating QPUs through a collection of platform and
instrument agnostic experiments, or \textit{routines}.

Thanks to its modular design, it offers routines with different abstraction
layers, from low-level characterization routines, including Rabi and Ramsey
experiments, to gate-level characterization algorithms~\cite{kr:2021:certification} such as randomized
benchmarking~\cite{eaz05:rb1,Knill_2008,llec:2007:rb3,dcel:2009:rb4,erowe:2022:rb_framework}.

With \Qibocal, it is possible to deploy various calibration and characterization
protocols and generate a comprehensive \texttt{HTML} report summarizing the results.
Alongside the report, \Qibocal also produces the new platform configuration containing
the fine-tuned parameters found.

When executing multiple experiments, these parameters can be updated at runtime,
allowing for complex routines with real time feedback.
This feature unlocks \Qibocal's potential to perform automatized hardware calibration,
which will be presented in a future manuscript in preparation \cite{qibocal_paper}.

\section{Quantum computing drivers}
\label{sec:drivers}

\texttt{Qibolab} provides a unified framework for controlling the different
electronics that are needed to operate a quantum computer.
To achieve this, we provide software abstractions and patterns that can be followed
by a laboratory in order to operate their self-hosted devices.
As a use case, we support drivers for multiple commercial instruments, which we use
to showcase the library and provide benchmarks in Sec.~\ref{sec:application}.
In the following sections we describe the software abstractions and supported
drivers in more detail.

\subsection{Software abstractions}
\label{sec:abstractions}

\begin{figure*}
  \center
  \includegraphics[width=0.7\textwidth]{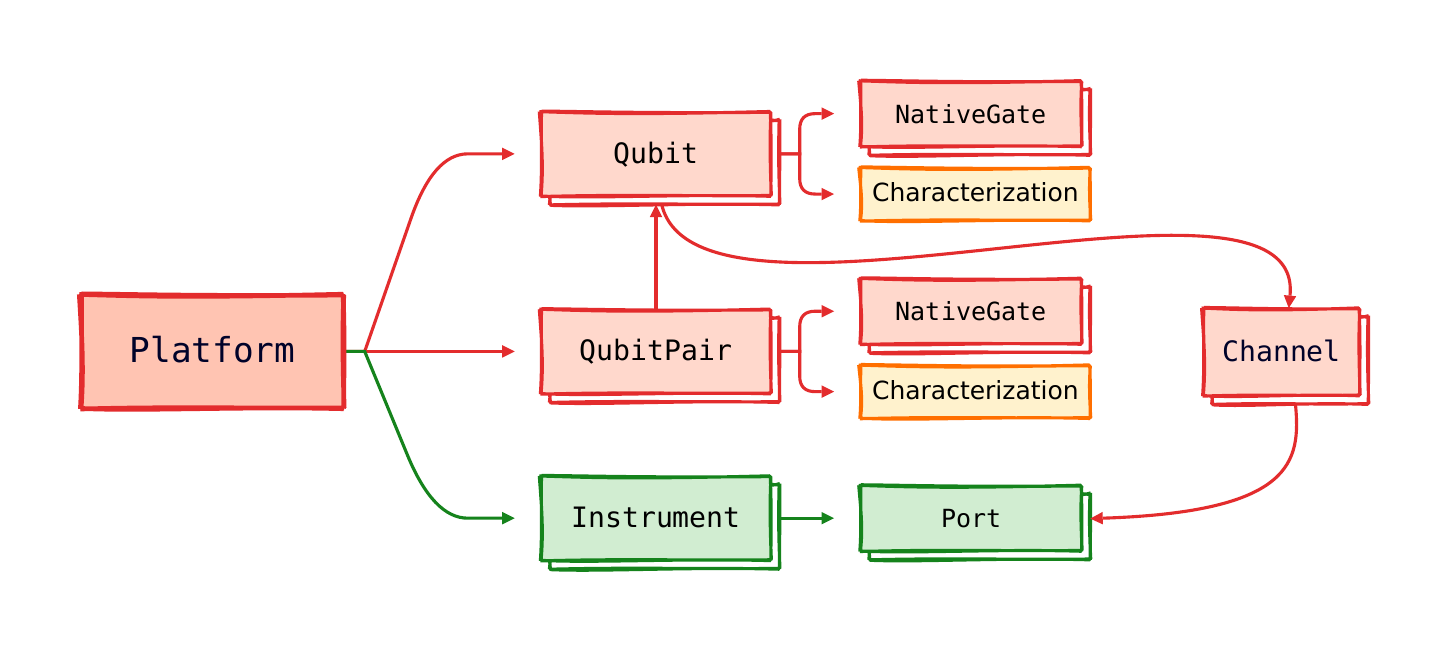}
  \caption{\label{fig:platform_object}Hierarchy of objects inside the \texttt{Platform}.}
\end{figure*}

\texttt{Qibolab} provides two main interface objects:
the \texttt{Pulse} object for defining arbitrary pulses to be played on qubits,
and the \texttt{Platform} which is used to execute these pulses on a specific QPU.

Pulses constitute the building blocks of programs that are executed on quantum hardware.
They can be used to read the state of a qubit, drive it to change its state,
or flux a qubit to change its resonance frequency and probe two-qubit interactions.
\Qibolab provides pulse objects for each of these operation modes and each
\texttt{Pulse} object holds information about the amplitude, frequency, phase, start
and duration of the pulse, which are required for the generation of physical pulses.
We also provide the functionality to generate waveforms of different shapes,
such as Rectangular, Gaussian or DRAG~\cite{drag-PhysRevLett.103.110501}.

Real experiments involve playing multiple pulses on different qubits.
In \texttt{Qibolab} pulses can be aggregated in a \texttt{PulseSequence}.
The \texttt{Pulse} API provides flexibility in scheduling such sequences by specifying
when each individual pulse starts in time and allowing overlapping pulses,
which are essential for features such as readout multiplexing~\cite{multiplex-PhysRevApplied.10.034040}.

Abstract sequences of pulses defined using the \texttt{Pulse} API can be deployed on
hardware using a \texttt{Platform}. This core \texttt{Qibolab} object is used to
orchestrate the different instruments for qubit control.
Each \texttt{Platform} instance corresponds to a specific quantum chip controlled
by a specific set of instruments. It allows users to execute a single sequence,
a batch of sequences, or perform a sweep, in which one or more pulse parameters
are being updated in real-time, within the control instrument.
Real-time sweeps or executing sequences in batches, significantly speeds up
qubit calibration and characterization procedures.

\texttt{Platform} is comprised of different objects as shown in Fig.~\ref{fig:platform_object}.
\texttt{Qubit} objects are representations of the physical qubits. They contain
information about physical parameters associated to a qubit that are measured
during calibration and characterization~\cite{calibration-10.1145/3529397, calibration-kelly2018physical},
such as coherence times $T_1$ and $T_2$, or the parameters of pulses and sequences
needed for single-qubit native gates.
Similarly, \texttt{QubitPair} objects contain information about the neighboring
pairs of qubits in a chip and the corresponding two-qubit native gates.
The topology of the chip is extracted from the available pairs and is
used by the transpiler presented in Sec.~\ref{sec:transpiler}.

\texttt{Platform} holds a collection of \texttt{Instrument} objects
which contain the low-level drivers for operating the laboratory
equipment. The abstract \texttt{Instrument} class contains the methods
one needs to implement when interfacing \texttt{Qibolab} to the libraries
provided  by the instrument's manufacturers, so that the instrument can be
used as part of a larger instrument setup compatible with
all functionalities provided by the \texttt{Qibo} framework.
\texttt{Controller} is a subclass used by instruments that have
arbitrary waveform generators and can play and acquire pulses.
\texttt{Qibolab} provides pre-coded driver implementations for
several commercial qubit control instruments, as described in
Sec.~\ref{sec:supported-drivers}.

Finally, \texttt{Channel} represents a connection from qubits to instruments.
Through the \texttt{Port} object it also implements an interface for controlling
instrument parameters. This connection is essential for playing pulses from the instrument
port that targets the desired qubit. It also provides a qubit-centric
interface for setting instrument parameters, which is useful in calibration routines.

To operate a real QPU, one needs to create a \texttt{Platform} that mirrors
the channel and instrument configuration of the lab,
following the example shown in Fig.~\ref{fig:qubit_scheme}.
The procedure is outlined in the following steps:
\begin{enumerate}
    \item instantiate \texttt{Instrument} objects for all instruments in the lab setup;
    \item create \texttt{Channel} objects for all connections between instruments and qubits,
    and map them to the corresponding instrument ports.
    Auxiliary instruments such as local oscillators can also be mapped to a \texttt{Channel};
    \item create a \texttt{Qubit} object for each qubit;
    \item assign all applicable channels (\texttt{readout}, \texttt{feedback}, \texttt{drive}, \texttt{flux}, \texttt{twpa}) to each \texttt{Qubit}.
    Note that a qubit may not have all of these channels and a channel may be shared
    among different qubits.
\end{enumerate}
Some parameters involved in this procedure, such as qubit-channel and channel-instrument connections
and instrument IP addresses are static, while others, such as the parameters of the pulses
that are implementing the native gates, change dynamically during qubit calibration.
It is important to distinguish these two categories and handle them separately in the code.
Static parameters are typically hard-coded in the \texttt{Platform} generation,
while dynamic parameters are loaded as external data.
More details on how a custom \texttt{Platform} can be written for
a specific lab setup can be found in the online documentation~\cite{qibolab-platform-tutorial}.
If the parameters of an existing platform are updated,
for example through a calibration routine, it is possible to dump the
new parameters on disk using serialization methods~\cite{qibolab-serialize}.
Parameters are uploaded to the respective devices using their specific API, which is
abstracted by the \texttt{Platform} interface.

Executing a program on the created \texttt{Platform} is also a multi-step process.
Users can write their programs using the \Qibolab \texttt{Pulse} API
or the \Qibo \texttt{Circuit} API. The former is commonly used for low-level applications
such as qubit calibration, while the latter is needed for executing quantum algorithms.
Execution of circuits involves additional steps. First, they are transpiled (see Sec.~\ref{sec:transpiler})
to new circuits that respect the QPU connectivity and native gates.
Secondly, native gates are compiled to pulses following a set of rules
which are held in the \texttt{Compiler} object of the \Qibolab backend.
Once a \texttt{PulseSequence} is available, either directly or from compilation of a circuit,
it can be deployed using a \texttt{Platform}. The \texttt{Platform} will send each pulse
to the appropriate instrument ports and acquire feedback associated to measurements.
This will be returned to the user according to the specified format.
The Qibolab port is an internal abstraction, that takes care of bridging the gap between
the Qibolab compiled \texttt{PulseSequence} and the output format to the specific input
and output defined by each device.
Available formats are classified shots (0 or 1), integrated and demodulated voltage signals,
or raw waveform signals. All formats can be obtained as single shots or averaged. More details on the different result formats can be found in
the online documentation~\cite{qibolab-results}.

\subsection{Supported drivers}
\label{sec:supported-drivers}

Version \texttt{0.1.0} of the \Qibolab package provides extensive support for various devices used in quantum hardware control.
Specifically, it supports devices developed by Qblox~\cite{Qblox}, Quantum Machines~\cite{quantum_machines}, Zurich Instruments~\cite{zurich_instruments}, as well as RFSoC (Radio Frequency System on Chip) FPGAs (Field Programmable Gate Arrays) supported by the \texttt{Qick} project~\cite{Stefanazzi2022} and by \texttt{Qibosoq}~\cite{rodolfo_carobene_2023_8126172}.
Each of these devices possesses distinct requirements and operational methods, necessitating meticulous attention to ensure seamless control through a unified interface.
In more detail:
\begin{description}

    \item[Qblox] the Qblox Instruments cluster~\cite{QbloxCluster} where  we  tested Qibolab is  composed  of  several modular devices controlled as one.
    The Qblox Cluster is the scalable 19" rack instrument that can be configured with a combination of up to 20 modules that can
    control and readout qubits over a wide frequency range (up to 18.5 GHz). Our setup to control 5 superconducting flux tunable qubits without coupler mediated interactions is composed by:

    \begin{description}
      \item[QRM-RF] two Qubit Readout Modules~\cite{QbloxQRMRF} with one input channel and one output channel in the radio-frequency regime and 2 digital markers.
      The module provides all necessary capabilities for qubit readout without external up or down conversion for signals in the range of 2-18.5GHz.

      \item[QCM-RF] three Qubit Control Modules~\cite{QbloxQCMRF} with two drive channels per module dedicated to the qubit control using parametrized pulses, that allows the user to control up to 5 qubits.

      \item[QCM] two QCM modules~\cite{QbloxQCM} to control the DC voltage applied to
      the flux channels of the qubits and generate the flux pulses needed to implement
      two-qubit gates. The dynamic output range of the DACs (digital-to-analog converters) of the Quantum Control Modules is 5 Vpp, the difference between the highest and the lowest
      voltage values in a AC signal, with a 1Gsps sampling rate.
    \end{description}

    The system synchronization of the signals between the modules is made by the Qblox Cluster using SYNQ~\cite{SYNC} protocols.
    The high-level interface for the devices comes from Qblox
    Instruments~\cite{QbloxInstruments} and Qcodes~\cite{QCODESsite} Python-based libraries, and the low-level communication with the sequencers is
    made using assembly code (Q1ASM)~\cite{Q1ASM}. This setup allows control of 5+ flux tunable superconducting qubits.

    \item[Quantum Machines] \Qibolab has been tested in controlling a cluster of nine OPX+ controllers~\cite{quantum_machines_opx},
    and communicate with an all-to-all connectivity to support fast feedback operations between any pair of controllers.
    The synchronization and clock distribution is handled by OPT devices.
    Each OPX+ controller has ten analogue output ports, ten digital output ports and two input ports, making the cluster capable of controlling 25+ flux tunable capacitively coupled qubits.
    The main disadvantage of our OPX+ controllers, compared to other instruments used in this work, is that the IQ mixing and upconversion are not taken care
    internally and there is small bandwidth for the intermediate frequency (400MHz) and output voltage (0.5V).
    Due to these limitations, additional external instruments including local oscillators, mixers and sometimes amplifiers are needed to successfully drive and flux qubits.
    The \Qibolab driver is controlling the whole cluster as a single instrument using the QUA library~\cite{qm_qua}.
    This library exposes many low-level operations to Python via an intuitive but rich
    set of commands, which expands beyond simple pulse scheduling and includes
    conditional logic, loops and complex mathematical operations.

    \item[Zurich Instruments] the Zurich Instruments cluster where we tested \Qibolab is composed of several modular devices controlled as one.
    \begin{description}
      \item[SHFQC] a single SHFQC~\cite{SHFQC}, that can control the drive and readout of up to 6 superconducting qubits connected to the same readout probe.
      The IQ mixing and upconversion are taken care of internally by using a proprietary cleaner signal upconversion and downconversion scheme~\cite{DSH_Scheme}
      with an instantaneous bandwidth around 1.2 GHz without the need for calibration. They also provide an output voltage of 2 Vpp.
      \item[HDAWG] two HDAWGs~\cite{HDAWG} to provide up to 8 DC-coupled single-ended analogue output channels each to control the flux pulses required to interact with qubits and couplers. Up to 5 Vpp output voltage.
      \item[PQSC] A single PQSC~\cite{PQSC}, to synchronize the previous devices via the low-latency, real-time communication link ZSync.
      The PQSC comes with 18 ZSync ports to distribute the system clock and synchronize the instruments.
      Furthermore, the links provide a bidirectional data interface to send qubit readout results to the PQSC for central processing and send trigger signals to the slave instruments for feedback.
    \end{description}

    The high-level interface for the devices comes from the Python-based LabOneQ library~\cite{LabOneQ}. This setup allows control of 5+ flux tunable superconducting qubits with tunable coupler-mediated interactions.

    \item[RFSoCs] the RFSoCs supported by \Qibolab currently include the RFSoC4x2~\cite{xilinxRFSoC4x2}, the ZCU111~\cite{xilinxZCU111}, and the ZCU216~\cite{xilinxZCU216} manufactured by Xilinx. These FPGAs possess a unique feature of offering direct RF synthesis capability up to $\approx9.8$ GHz. This simplifies the experimental setup by eliminating the need for additional local oscillators and IQ mixers. To interact with the \texttt{Qick} firmware, the driver relies on a server that runs on board called \texttt{Qibosoq}. Both the \texttt{Qibosoq} server and the \texttt{Qick} firmware are open source, reducing costs for setting up a new laboratory. However, it is important to note that these boards have limitations in terms of the number of qubits they can control, can be challenging to synchronize in multi-board setups and, in general, the software supports less features than other devices.
\end{description}

In addition to the devices responsible for synthesizing pulses to control the qubits and acquiring signals for measurements, a comprehensive quantum control system relies on additional devices.
Among these, local oscillators play a crucial role in up and down converting microwave signals for some of our devices and pumping the TWPAs.
Integrating local oscillators within the same framework is essential since they need to be calibrated and turned on and off during the control process.
\Qibolab facilitates seamless integration of these devices and includes drivers for \textit{Erasynth} and \textit{Rohde\&Schwarz} local oscillators in version \texttt{0.1.0}.

\begin{table}
  \scriptsize
    \begin{tabular}{lcc}
    \hline \hline
	    \textbf{Device}     & \textbf{Firmware}          & \textbf{Software}       \\ \hline
	    Qblox      & 0.4.0 & qblox-instruments 0.9.0                 \\
    	QM         & QOP213           & qm-qua 1.1.1      \\
    	Zurich     & Latest (July 2023)\footnote{See appendix [\ref*{ZurichFirmware}]}                 &   LabOneQ 2.7.0              \\
	    RFSoCs     & Qick 0.2.135     & Qibosoq 0.0.3     \\ \hline
    	Erasynth++ & -                & -     \\
	\cell{R\&S SGS100A}    & -                & QCoDeS 0.37.0    \\
  \hline \hline
    \end{tabular}
\caption{Outline of the supported devices, along with firmware/software version currently supported.}
\label{tab:drivers}
\end{table}

An outline of the supported instruments is presented in Table~\ref{tab:drivers},
while in Table~\ref{tab:drivers-features} we present an overview of the primary
features supported by the drivers included with \Qibolab version \texttt{0.1.0}.
It is important to note that while some limitations and missing features are
currently present, they are not necessarily inherent to the devices themselves
and will be addressed in future versions of \Qibolab.

\begin{table}
  \scriptsize
	\begin{tabular}{lcccc}
  \hline \hline
		\textbf{Feature}              & \textbf{RFSoCs}    & \textbf{Qblox}    & \textbf{QM}        & \textbf{Zhinst}   \\ \hline
		Arbitrary pulse sequences     & \usym{1F5F8}  & \usym{1F5F8} & \usym{1F5F8}  & \usym{1F5F8} \\
		Arbitrary waveforms           & \usym{1F5F8}  & \usym{1F5F8} & \usym{1F5F8}  & ~\usym{1F5F8}\footnote{Sweeper capabilities may be reduced by using arbitrary pulses instead of driver defined ones.}      \\
		Multiplexed readout           & \usym{1F5F8}  & \usym{1F5F8} & \usym{1F5F8}  & \usym{1F5F8} \\
		Hardware classification       & \usym{2613}  & \usym{1F5F8} & \usym{1F5F8}  & \usym{1F5F8} \\
		Fast reset                    & \usym{1F4BB}   & \usym{1F4BB} & \usym{1F4BB}   & \usym{1F4BB}  \\
		Device simulation             & \usym{2613}  & \usym{2613} & \usym{1F5F8}  & \usym{1F4BB}  \\
		RTS frequency                 & ~\usym{1F5F8}\footnote{RTS on the frequency of readout pulses not supported.}    & \usym{1F5F8} & \usym{1F5F8}  & \usym{1F5F8}      \\
		RTS amplitude                 & \usym{1F5F8}  & \usym{1F5F8} & \usym{1F5F8}  & \usym{1F5F8} \\
		RTS duration                  & \usym{2613}  & \usym{1F5F8} & \usym{1F5F8}  & \usym{1F5F8} \\
		RTS start                     & \usym{1F5F8}  & \usym{1F5F8} & \usym{1F5F8}  & \usym{1F5F8} \\
		RTS relative phase            & \usym{1F5F8}  & \usym{1F5F8} & \usym{1F5F8}  & \usym{1F5F8} \\
		RTS 2D any combination        & \usym{1F5F8}  & \usym{1F5F8} & \usym{1F5F8}  & \usym{1F5F8} \\
		Sequence unrolling            & \usym{1F4BB}   & \usym{1F4BB} &\usym{1F4BB}  & \usym{1F4BB} \\
		Hardware averaging            & \usym{1F5F8}  & \usym{1F5F8} & \usym{1F5F8} & \usym{1F5F8} \\
		Singleshot (No Averaging)     & \usym{1F5F8}  & \usym{1F5F8} & \usym{1F5F8} & \usym{1F5F8} \\
		Integrated acquisition        & \usym{1F5F8}  & \usym{1F5F8} & \usym{1F5F8} & \usym{1F5F8} \\
		Classified acquisition        & \usym{1F5F8}  & \usym{1F5F8} & \usym{1F5F8}  & \usym{1F5F8} \\
		Raw waveform acquisition      & \usym{1F5F8}  & \usym{1F5F8} & \usym{1F5F8}  & \usym{1F5F8} \\
    \hline \hline
	\end{tabular}
	\caption{Features or limitations of the main drivers supported by \Qibolab \texttt{0.1.0}.
	The features denoted by `` \protect\usym{1F5F8} " are supported, `` \protect\usym{2613} "
  means not supported and `` \protect\usym{1F4BB} " under development.}
	\label{tab:drivers-features}
\end{table}

The following is a description of the features presented in Table~\ref{tab:drivers-features}.
\begin{description}
    \item[Arbitrary pulse sequences] the capability of executing arbitrary pulse sequences defined in \Qibolab, which is a fundamental requirement of a driver. This feature is not related to the execution of pulses with arbitrary \textit{waveform shapes}.
    \item[Arbitrary waveforms] the capability of executing pulse waveforms of arbitrary shape. For drivers that do not support this feature, rectangular, Gaussian and DRAG waveforms can still be synthesized.
    \item[Multiplexed readout] allows playing and acquiring multiple multiplexed pulses through the same line. It is particularly useful for multi-qubit chips where the readout line is commonly shared among multiple qubits.
    \item[Hardware classification] the capability of doing single shot measurement classification \textit{during the execution} of a pulse sequence.
    \item[Fast reset] the capability of actively resetting the state of a qubit to zero after a measurement. This feature requires hardware classification and enables faster executions of repeated pulse sequences.
    \item[Device simulation] the possibility of simulating in advance the pulses to be executed, without directly using quantum hardware.
    \item[RTS frequency] RTS (\textit{Real Time Sweeper}) refers to the capability of executing a pulse sequence multiple times with different values of, in this case, the frequency of a pulse. This feature facilitates faster qubit characterization and experiments.
    \item[RTS amplitude] real-time sweeping of the amplitude of a pulse.
    \item[RTS duration] real-time sweeping of the duration of a pulse.
    \item[RTS start] real-time sweeping of the start time of a pulse.
    \item[RTS relative phase] real-time sweeping of the relative phase of a pulse.
    \item[RTS 2D] the capability of combining two RTS scans on different parameters.
    \item[Sequence unrolling] the capability of unrolling several smaller subsequences into a longer single sequence as in loop unrolling. It aims to decrease the overall time spent on compilation and communication steps by reducing its amount from once every subsequence to once every unrolled sequence.
    \item[Hardware averaging] the capability of repeating the same experiment multiple times and obtain, directly from the device, averaged results.
    \item[Singleshot (No Averaging)] the capability of obtaining from the devices all the non-averaged results.
    \item[Integrated acquisition] the capability of acquiring complex signals~\cite{Naidu2003} with "in-phase" and "quadrature" (IQ) components demodulated and integrated for the measuring time.
    \item[Classified acquisition] the capability of performing 0-1 state classification after the integrated acquisition.
    \item[Raw waveform acquisition] the capability of acquiring non-integrated IQ waveform values.
\end{description}

\subsection{Transpiler}\label{sec:transpiler}
Logical quantum circuits for quantum algorithms are hardware agnostic. Usually an all-to-all qubit connectivity is assumed while most current hardware only allows the execution of two-qubit gates on a restricted subset of qubit pairs.
Moreover, quantum devices are restricted to executing a subset of gates, referred to as \emph{native}~\cite{Barenco_1995}.
This means that, in order to execute circuits on a real quantum chip, they must be transformed into an equivalent,
hardware specific, circuit. The transformation of the circuit is carried out by the transpiler
through the resolution of two key steps: connectivity matching~\cite{ito2023algorithmic} and native gates decomposition~\cite{9894863}.
In order to execute a gate between two qubits that are not directly connected SWAP gates~\cite{article_zhu} are required.
This procedure is called \emph{routing}. As on NISQ devices two-qubit gates are a large source of noise, this procedure generates an overall noisier circuit. Therefore, the goal of an efficient routing algorithm is to minimize the number of SWAP gates introduced. An important step to ease the connectivity problem, is finding an optimal initial mapping
between logical and physical qubits. This step is called \emph{placement}.
The native gates decomposition in the transpiling procedure is performed by the \emph{unroller}. An optimal decomposition uses the least amount of two-qubit native gates. It is also possible to reduce the number of gates
of the resulting circuit by exploiting commutation relations~\cite{itoko2019optimization}, KAK decomposition~\cite{Vidal_2004}
or machine learning techniques~\cite{fosel2021quantum}.

\Qibolab implements a built-in transpiler with customizable options for each step. The main algorithms that can be used at each
transpiler step are reported below with a short description.
The initial placement can be found with one of the following procedures:
\begin{itemize}[nosep]
\item Trivial: logical-physical qubit mapping is an identity.
\item Custom: custom logical-physical qubit mapping.
\item Random greedy: the best mapping is found within a set of random layouts based on a greedy policy.
\item Subgraph isomorphism: the initial mapping is the one that guarantees the execution of most gates at the beginning of
the circuit without introducing any SWAP.
\item Reverse traversal: this technique uses one or more reverse routing passes to find an optimal mapping by starting from
a trivial layout~\cite{li2019tackling}.
\end{itemize}
The routing problem can be solved with the following algorithms:
\begin{itemize}[nosep]
\item Shortest paths: when unconnected logical qubits have to interact, they are moved on the chip on the shortest path
connecting them. When multiple shortest paths are present, the one that also matches the largest number of the following two-qubit gates is chosen.
\item SABRE~\cite{li2019tackling}: this heuristic routing technique uses a customizable cost function to add SWAP gates that reduce the distance between unconnected  qubits involved in two-qubit gates.
\end{itemize}
\Qibolab unroller applies recursively a set of hard-coded gates decompositions in order to translate any gate into single and two-qubit native gates. Single qubit gates are translated into U3, RX, RZ, X and Z gates. It is possible to fuse multiple single qubit gates acting on the same qubit into a single U3 gate. For the two-qubit native gates it is possible to use CZ and/or iSWAP. When both CZ and iSWAP gates are available the chosen decomposition is the one that minimizes the use of two-qubit gates.

\begin{figure}
  \center
  \includegraphics[width=1\linewidth]{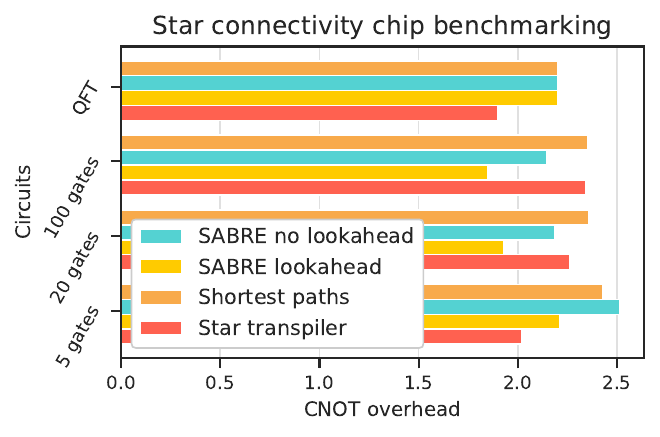}
  \caption{Benchmark of the performance of the built-in \Qibolab transpilers (routing pass) evaluated considering the CNOT overhead.
  Circuits of different types have been considered in order to test the transpilers on both structured and unstructured circuits. In
  particular we have considered a five qubits QFT circuit and random circuits with 5, 20 and 100 gates.
  The results for random circuits have been averaged over 50 circuits. }
  \label{fig:transpiler}
\end{figure}

The benchmarking of a full transpiling pipeline can be complex, as the results may vary in different chip architectures and a trade-off between performance and execution time needs to be taken into account.
We remand the general benchmarking problem to the specific literature~\cite{kharkov2022arline} and we focus on a more specific use-case.
Fig.~\ref{fig:transpiler} reports the performance of the routing pass algorithms implemented in \Qibolab on a five qubit chip with a star connectivity.
In this kind of chip five qubits are arranged with a central qubit connected to all the remaining four qubits.
The algorithm performance has been evaluated as the CNOT overhead. That is the number of CNOT gates on the routed circuit divided by the number of CNOT
gates in the original circuit.
The algorithm performance has been tested on five qubit circuits composed of 10, 20 and 100 CNOT gates, taking an average over 50 random circuits.
Moreover, we have tested the algorithms on a structured circuit: the five qubits QFT. The SABRE algorithm has been tested with and without the lookahead.
The results have been compared with transpiler designed for that connectivity (star transpiler) that swaps the central qubit in the chip based on the successive gate.
All the routing algorithms have been tested starting from an initial trivial layout except for the star transpiler that has a built-it placer, this explains
the better performance of this algorithm on short circuits.
Fig.~\ref{fig:transpiler} shows that SABRE with a lookahead reaches the best performance on the star connectivity chip.
The execution time on this simple case is not significant as all algorithms perform in a fraction of second even for the longer circuits.
However, other tests has shown that the scalability of SABRE is better than shortest paths as the number of possible shortest paths increases drastically with the
number of qubits in highly connected chips. In summary, \Qibolab transpiler shows good performance in making abstract quantum circuits executable on small NISQ devices.
In the future we aim at developing new efficient and scalable algorithms for the next generation quantum chips with a high number of qubits.

\section{Application results}
\label{sec:application}

\subsection{Cross-platform benchmark}\label{sec:benchmark}

\begin{figure*}
  \center
    \includegraphics[width=0.99\textwidth]{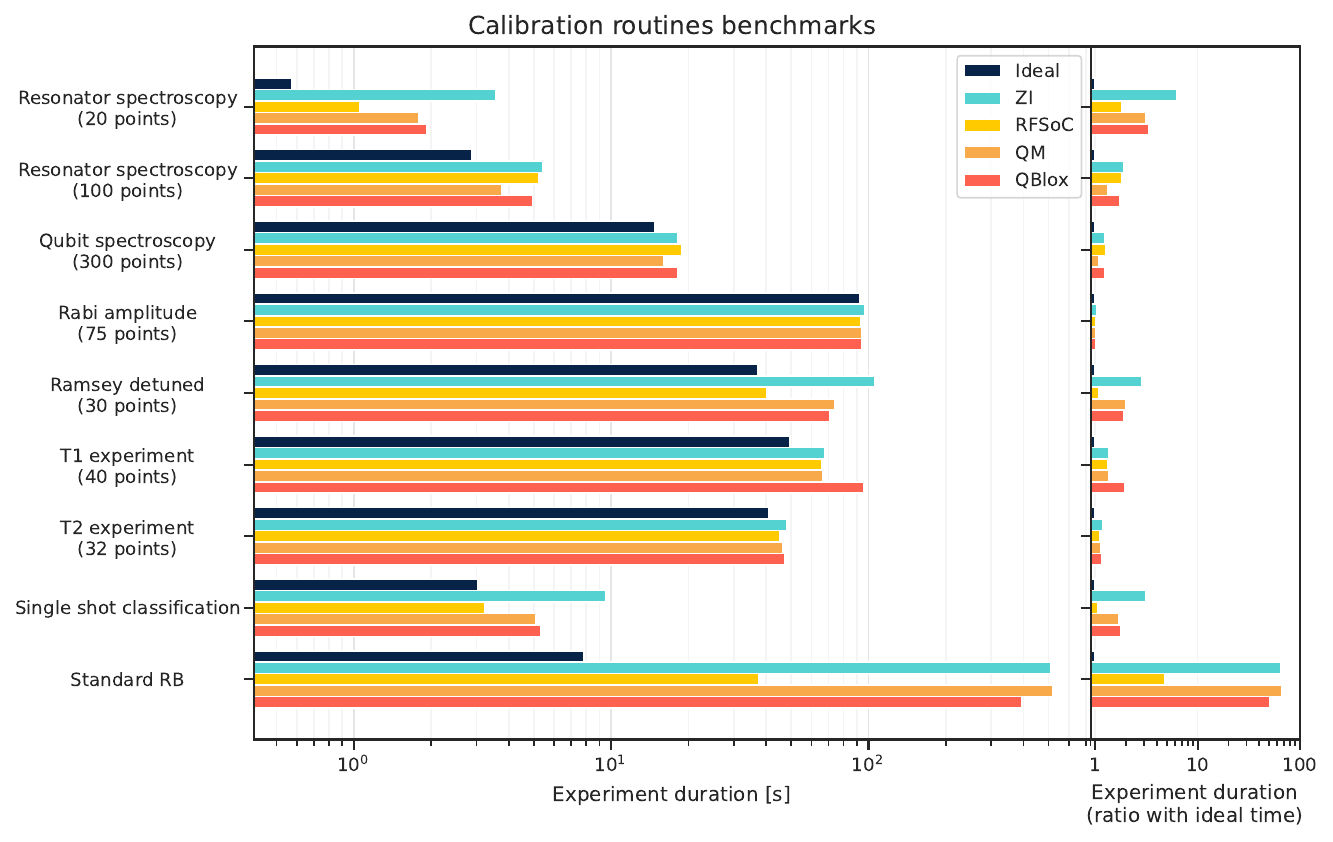}
    \caption{Execution time of different qubit calibration routines on various electronics. On the left side we show the absolute times in seconds for each experiment. The ideal time (black bar) shows the minimum time the qubit needs to be affected in each experiment. On the right side we calculate the ratio between actual execution time and ideal time. Real-time sweepers are used, if supported by the control device, in all cases except the \textit{Ramsey detuned} and \textit{Standard RB} experiments.}
  \label{fig:speed_benchmark}
\end{figure*}

\begin{figure*}
  \center
  \begin{tabular}{ccc}
      \includegraphics[width=0.31\textwidth]{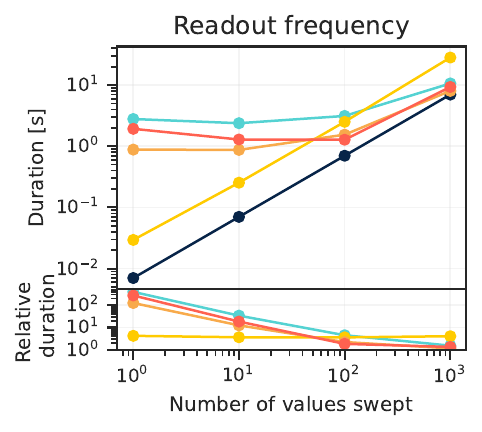}
      &
      \includegraphics[width=0.31\textwidth]{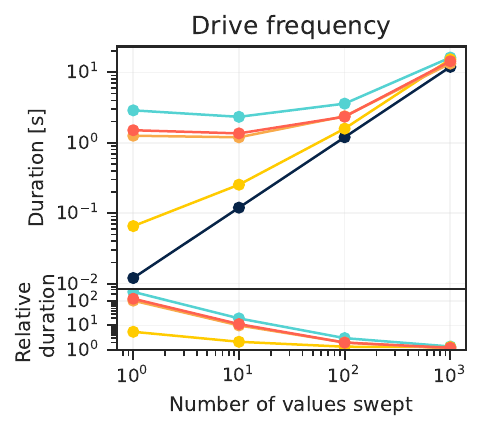}
      &
      \includegraphics[width=0.31\textwidth]{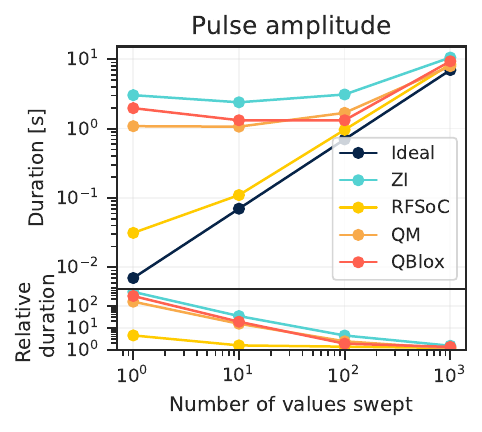}
      \\
      \includegraphics[width=0.31\textwidth]{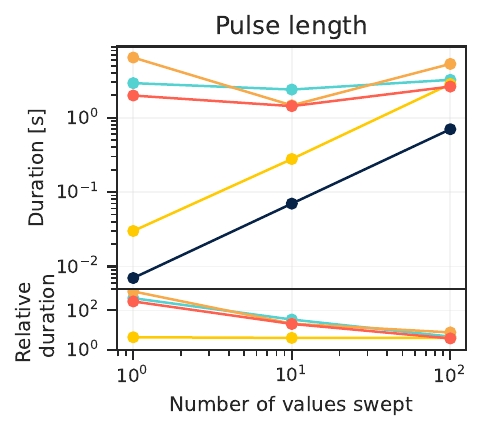}
      &
      \includegraphics[width=0.31\textwidth]{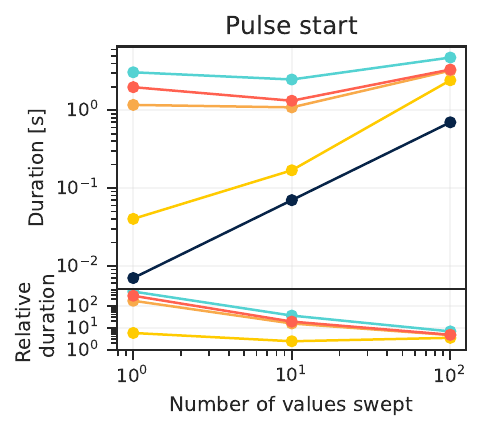}
      &
      \includegraphics[width=0.31\textwidth]{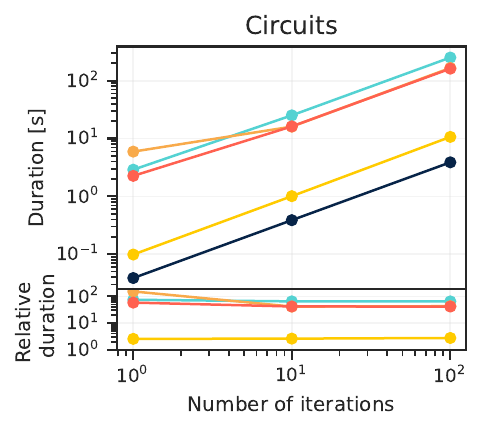}
  \end{tabular}
      \caption{Scaling of execution time as a function of the number of points in a sweep. Bottom plots show the ratio between real execution on different instruments and minimum ideal time. Real-time sweepers are used in all cases, except the last \textit{Circuits} plot where we use the standard RB experiment to generate a given number of random circuits to execute.}
      \label{fig:benchmark_scaling}
  \end{figure*}
In this section, we present the results of a speed benchmark conducted using \Qibolab. The benchmark involved various experiments deployed on the different control devices currently supported by the drivers implemented in \Qibolab.

By utilizing \Qibolab, assessing the performance and efficiency of each control device becomes a straightforward process, since all devices are exposed through the same interface.
The results obtained not only offer valuable insights into the speed of the different instruments, offering data that can help researchers and developers to make informed decisions, but also demonstrate the comprehensive support these devices receive within \Qibolab.

The experiments chosen for this benchmark represent the minimal set of routines required for the calibration of a single qubit. They also offer a view of the different execution modes supported by \Qibolab: in particular the \textit{Single shot classification} experiment executes fixed pulses sequences, while the \textit{Spectroscopies} perform different sweeps over pulse parameters.

In Fig.~\ref{fig:speed_benchmark} we present a comparison of execution times
for different qubit calibration routines executed using different electronics.
Additional details for each routine are provided in the Appendix~\ref{appendix:benchmark}.
The black bar in this plot provides the $\mathrm{ideal}$ time required for each
routine, which, in most cases, is calculated as
\begin{equation}
    \mathrm{ideal} =n_{\mathrm{shots}} \sum_i (T_{\mathrm{sequence}, i} + T_\mathrm{relaxation})
\end{equation}
where $T_{\mathrm{sequence}, i}$ is the duration of the whole pulse sequence
in the $i$-th point of the sweep, $T_{\mathrm{relaxation}}$ the time we wait
for the qubit to relax to its ground state between experiments,
$n_{\mathrm{shots}}$ the number of shots in each experiment and
the sum runs over all points in the sweep.
The $\mathrm{ideal}$ time denotes how long the qubit is really used during
an experiment and provides the baseline for our benchmark.
Real executions, shown with a different color for each instrument setup,
are longer than $\mathrm{ideal}$, due to overhead coming from compilations and
communication to the instruments. After profiling the code, we observe that the overhead
coming from the \Qibolab backend, $T_{\mathrm{qibo}}$, is negligible compared to that of the control instruments, $T_{\mathrm{inst}}$.
Therefore, we can approximate the real execution time as
\begin{equation}
  \mathrm{real} = T_{\mathrm{qibo}} + T_{\mathrm{inst}} + \mathrm{ideal} \eqsim T_{\mathrm{inst}} + \mathrm{ideal}
\end{equation}

There is a decisive factor regarding the performance of routines that involve sweeps.
That is, whether the sweeps run in real-time in the processors embedded in
the control electronics or the host computer. The latter approach requires
a greater number of communication steps between control electronics and host and typically the programs
need to be recompiled multiple times resulting in significant overhead.
This can be seen in the \textit{Ramsey detuned} and
\textit{standard Randomized Benchmarking (RB)} experiments,
for which real time sweepers have not been implemented yet,
resulting to a significant overhead over the ideal time.
Randomized Benchmarking experiments, unlike the rest of routines used here,
involve playing multiple random sequences instead of sweeping parameters and
their performance is expected to increase when sequence unrolling will be implemented.

The second point affecting performance is the communication with the host computer.
This usually involves two steps, the actual communication via network (ethernet)
and a compilation step happening on the instrument side.
We observe that RFSoC boards controlled using \texttt{Qibosoq} have an advantage
in this, particularly from \textit{Ramsey detuned} and \textit{Single shot classification} where real-time sweepers are not used.
This advantage may be due to the simplicity of our RFSoC configuration,
which consists of a single board, in contrast to the other systems which are part
of clusters with more controllers. More investigation is needed to confirm
this point. As expected, the rest of electronics behave similarly in all
performed benchmarks.

In Fig.~\ref{fig:benchmark_scaling} we demonstrate how execution time of
different sweeps scales with the number of points used in the sweep. Similarly
to above, we see that RFSoC is faster for short sweeps, due to smaller
communication and compilation overhead, however the difference diminishes when
we cross 100 points. Other instruments show similar behavior in most cases.
\texttt{Qick} does not support real-time sweeping of readout frequency and pulse
length, therefore these sweepers are slower when compared to other instruments
for more than 100 points. Real-time sweepers are used in all cases presented
in this plot, except in \textit{Circuits}.
We are currently implementing sequence unrolling methods, which will allow
executing batches of circuits, reducing the communication overhead and thus
improving runtime.

The code used for all benchmarks presented in this section is provided in a public repository~\cite{qibolab-benchmarks}.

\subsection{Standard randomized benchmarking}

\begin{figure}
  \center
  \includegraphics[width=1\linewidth]{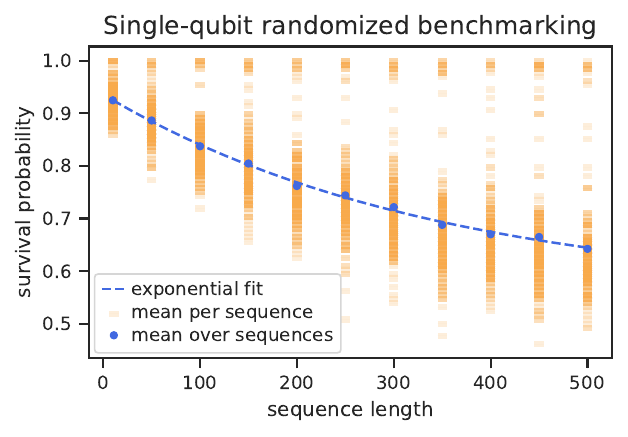}
  \caption{\label{fig:std_rb}
  Results of single-qubit randomized benchmarking experiment implemented in \Qibo\ and executed with \Qibocal\,
  on a $5$-qubit IQM chip controlled by \Qibolab's Zurich Instruments drivers.
  Average relative frequency (survival probability) of classified $0$s of $128$ single-shot measurements (orange) for random sequences of single-qubit Clifford gates of different length and mean over $256$ random sequences (blue).
  The exponential fit is described by $m \mapsto 0.38(2)\cdot 0.9971(3)^m + 0.55(2)$. This corresponds to an average gate fidelity of $0.9986(2)$ and a $\pi/2$-pulse fidelity of $0.9992(1)$.
  Errors are the standard deviation of $1000$ `semi-parametric' bootstrapping samples (of binomial random variables with parameter drawn from the empirically observed distribution of relative frequencies).
  }

\end{figure}

The commonly used technique for
assessing the accuracy of single qubit gate implementations is standard randomized benchmarking (RB) \cite{eaz05:rb1,Knill_2008,llec:2007:rb3,dcel:2009:rb4,erowe:2022:rb_framework} with the Clifford group (see, e.g., Ref.~\cite{kr:2021:certification} for a review).
The RB protocol performs random sequences of Clifford unitaries of different lengths on a single qubit.
Every sequence is concluded with the unitary gate that restores the initial state before measuring the qubit.
In the absence of imperfections, the measurement, thus, is expected to be classified as $0$ (the initial state) with probability $1$ independent of the sequence or its length.
Single qubit gate fidelities are defined as functions of the decay parameter of this average \emph{survival probability} with the sequence length.

RB allows us to holistically test the entire software stack together with the quantum hardware.
We define the RB protocol with the \texttt{Circuit} API of \Qibo, using U3, RX, RY and RZ gates.
Executing them with the \Qibolab\ backend involves transpilation to native gates and compilation to \texttt{PulseSequence} objects that are then executed by a \texttt{Platform}.
An example of an RB experiment on a $5$-qubit IQM chip controlled using \Qibolab's Zurich Instruments drivers is depicted in Fig.~\ref{fig:std_rb}.

\subsection{CHSH Experiment}

A quantum software solution should allow for the development and deployment of quantum experiments at different levels of abstraction and complexity. In order to showcase this, we prepare an experiment to measure the CHSH inequality \cite{clauser1969chsh} between two qubits by building the circuit using three distinct methods allowed by \Qibo and \Qibolab. Namely, one can build experiments by directly accessing the arbitrary pulse sequences of \Qibolab, use the native gate interactions through \Qibo, or use logical gate operations and rely on the \emph{transpiler} for the decomposition. In addition, we incorporate a layer of Readout Error Mitigation \cite{van_den_Berg_2022} that is executed on hardware before the experiment takes place. This type of control is only possible with framework aware of all layers of abstraction such as \Qibo.

The CHSH inequality was originally conceived in order to disprove a local hidden-variable description of quantum mechanics and used to prove Bell's theorem \cite{bell1964epr}. The protocol consists of preparing a maximally entangled two-qubit state, and performing a simultaneous measurement on both qubits, with two possible measurement settings. Crucially, a qubit can be measured in two perpendicular basis (e.g.~$X$, $Z$), and the measurement settings of both qubits have a relative angle $\theta$. Then, the combination of the resulting expectation values
\begin{equation}\label{eq:chsh}
    S = E(a, b) - E(a, b') + E(a', b) + E(a', b'),
\end{equation}
should be $\big|S\big|\leq2$ if there is a local hidden-variable theory of quantum mechanics, but go beyond, up to $2\sqrt{2}$, if not.

\begin{figure}
  \center
  \includegraphics[width=1\linewidth]{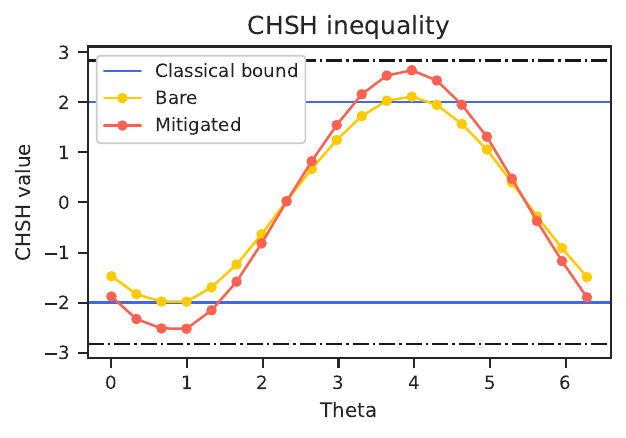}
  \caption{Results of bare (yellow) and mitigated (red) CHSH values from an experiment
  on two qubits on a 5-qubit QuantWare chip controlled via \Qibolab's Qblox drivers.
  Readout error mitigation significantly enhances the results of the CHSH inequality,
  bringing it past the classical bound (blue line). The initial entangled state prepared
  for this experiment is $(\ket{01}-\ket{10})/\sqrt{2}$. The significant improvement
  produced by readout error mitigation hints that readout error dominates in the
  deterioration of the experimental results.}
  \label{fig:chsh}
\end{figure}

In this context, the CHSH experiment is used as validation of both the specifications of the machine and control electronics. Far from disproving local realism, we use this procedure to verify that the control of our chip is precise enough to violate the classical bound. We show in Fig. \ref{fig:chsh} the results of a CHSH experiment on two connected qubits of a 5-qubit QuantWare chip controlled using \Qibolab's Qblox drivers for different angles of the measurement setting. The bare values of the CHSH barely cross above the classical bound. However, when readout error mitigation is applied, the values confidently cross above $2$. We infer from this figure that the most destructive source of error was in the readout of the measurement pulse, rather than on the control of the two-qubit gate pulse.

\subsection{Full-stack quantum machine learning}

\begin{figure}
  \center
  \includegraphics[width=1\linewidth]{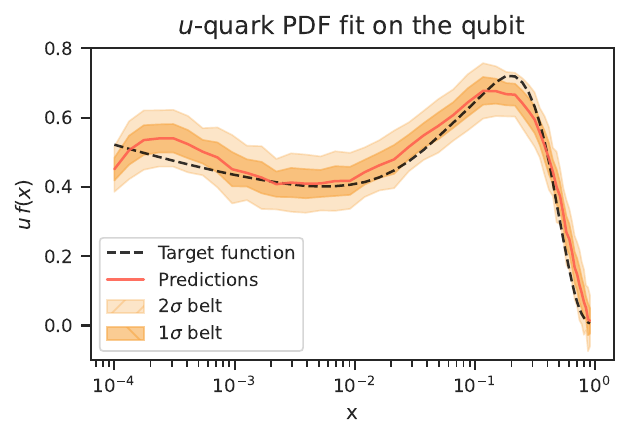}
  \caption{Estimates of $N_{\rm data}=50$ points of the $u$-quark PDF
  using the 1-qubit device controlled by the RFSoC. The target values (black line) are
  compared with the estimates obtained with the qubit. The solid orange line and the
  confidence intervals are calculated by repeating $N_{\rm runs}=50$ times the
  estimations with the trained model and then calculating means and standard deviation
  of the mean of the $N_{\rm runs}$ predictions. In particular, the two confidence
  intervals are computed using $1\sigma$ and $2 \sigma$ errors.}
  \label{fig:qml_example}
\end{figure}

Developing QML algorithms~\cite{Schuld_2014,
Biamonte_2017, Mitarai_2018} is particularly challenging in the NISQ
era~\cite{Preskill_2018}. Noise and long execution times are two of the most
limiting problems while deploying a Variational Quantum
Algorithm~\cite{Cerezo_2021, Wang_2021} on a real quantum device. In this
context, it is relevant to study how the different levels of computation, from
the high-level coding of the algorithm to the low-level deployment on the real
qubits, impact the results obtained on simple regression or classification
tasks. For this reason, \Qibo has become the perfect environment to study both
hybdrid~\cite{P_rez_Salinas_2021, robbiati2023determining, Bravo_Prieto_2022, particles6010016}
and full-stack~\cite{robbiati2022quantum} QML algorithms.

We define a QML model by building a Variational Quantum Circuit (VQC), in whose
rotational gates we encode the $u$-quark Parton Distribution Function (PDF) data picked
up from the NNPDF4.0~\cite{Ball_2022} PDF grid. In particular, we consider as input data
$N_{\rm data}=50$ values of the momentum fraction $x$, sampled logarithmically from
the range $[0,1]$.
We use the model presented in~\cite{P_rez_Salinas_2021}, in which
the embedding of the $x$ values is implemented following a data re-uploading ansatz~\cite{P_rez_Salinas_2020}.

The optimization strategy is then
implemented by minimizing a target Mean-Squared Error loss function with respect
to the model's parameters. We select a hardware-compatible Adam~\cite{kingma2017adam}
optimizer, in which we calculate the derivatives of the circuit using the
Parameter Shift Rule~\cite{Mitarai_2018, Schuld_2019, robbiati2022quantum}.
Once obtained the optimized parameters vector $\bm{\theta}_{\rm best}$,
we inject them into the circuit and repeat the predictions $N_{\rm runs}=50$ times.
With the mean and the standard deviation $\sigma$ of these evaluations we calculate
our final estimates and their errors. Finally, we quantify the accuracy of the model
by computing the following test statistics:
\begin{equation}
\text{MSE} = \frac{1}{N_{\rm data}} \sum_{j=1}^{N_{\rm data}} \bigl( y_{j, \rm est} - y_{j, \rm target} \bigr)^2,
\label{eq:test_statistic}
\end{equation}
where $y_{j, \rm target}$ is the target PDF value provided by NNPDF4.0 and $y_{j, \rm est}$ is
the mean of the $N_{\rm runs}$ predicted values for a fixed data $x_j$.

We perform an initial training in exact simulation using \Qibo, followed by
$N_{\rm epochs}=60$ stochastic Adam descent iterations on a single superconducting
qubit controlled by RFSoC via \Qibosoq~\cite{rodolfo_carobene_2023_8126172}.
After completing the training, those corresponding to the epoch in which
we recorded the lowest loss function value are chosen as the final parameters.
Each prediction during the gradient descent is obtained by executing the circuit
$N_{\rm nshots}=500$ times and we set a learning rate equal to $\eta=0.1$.
Adam's parameters are set to be $\beta_1=0.85$, $\beta_2=0.99$
and $\varepsilon=10^{-8}$.

In Fig.~\ref{fig:qml_example} we show the obtained results after repeating the
PDF predictions $N_{\rm runs}=50$ times for each data: the solid orange line is drawn
using the means of the predictions $\{y_{j,\rm est}\}_{j=1}^{N_{\rm data}}$,
while the two confidence intervals are obtained one and two standard deviations from the means.
The test statistic value  presented in Eq.~\eqref{eq:test_statistic} and calculated
with our predictions is $\text{MSE}=0.0021$.
These results show the entire ecosystem can be used to successfully fit the target
function even without any error mitigation technique.

\section{Outlook}
\label{sec:outlook}

In this paper we extend the \Qibo quantum computing middleware framework by
introducing \Qibolab, an open-source software library for quantum hardware
control.
\Qibo is designed as a full-stack software framework which provides primitives
to define circuit-based quantum algorithms through custom backends,
\textit{i.e.}~dedicated plugin software libraries which deploy algorithms on
specific hardware. The release of \Qibolab unlocks \Qibo's potential to execute
quantum algorithms on hardware platforms and therefore grant to research
institutions and laboratories the possibility to operate self-hosted quantum
hardware platforms easily.

We have described the current status of the project structure with the major
features implemented in release \texttt{0.1.0}. The software abstractions,
supported drivers and transpiler are at the stage of allowing applications
related to cross-platform control instrument performance benchmarks through
arbitrary pulse control and physics experiments based on the quantum circuit
representation.

Furthermore, we have demonstrated successfully three practical-cases in which
\Qibolab could be useful for quantum technology research: randomized
benchmarking, validation algorithms for qubit entanglement (CHSH experiment) and
quantum machine learning applications. Therefore, circuit-based models available
in \Qibo can be deployed seamlessly on quantum hardware through \Qibolab.

In the future releases of \Qibolab, we plan to extend its capabilities by
interfacing new drivers from more commercial and open-source control system
vendors. Thanks to the design of the library, we have the possibility to adapt
and scale the API for new electronics including large-scale systems for
real-time acquisition and error correction.
On the other hand, in this paper we have focused on superconducting chips due to
its availability in our affiliated institution labs, however we plan to extend
\Qibolab to other quantum technologies such as trapped ions, neutral atoms and
photonics among others. In fact, there are multiple software
similarities among these technologies, \textit{e.g.}~for trapped ions
we can already define in \Qibolab a custom platform which allocates
the relevant pulse sequence to modulate optical lasers with its native gates
representation for unitary gate preparation. We plan to have access to this and
other quantum hardware technologies in the next years through research
collaborations and extend \Qibolab accordingly.
Finally, we believe that with the inclusion of \Qibolab, \Qibo has grown into a powerful tool for the
quantum computing community, by reducing the effort of software development for researchers
in simulation, hardware calibration and operation.

The code implementing the \Qibolab module is available at:
\begin{center}
  \href{https://github.com/qiboteam/qibolab}{\color{blue}\texttt{https://github.com/qiboteam/qibolab}}.
\end{center}

\acknowledgments This project is supported by TII's Quantum Research Center. The
authors thank all \Qibo contributors for helpful discussion. M.R.~is supported
by CERN's Quantum Technology Initiative (QTI) through the Doctoral Student
Program. M.R.~and S.C.~thank CERN TH hospitality during the elaboration of this
manuscript.

\bibliographystyle{apsrev4-2}
\bibliography{references}

\section{Appendix}

\subsection{Zurich Instruments firmware} \label{ZurichFirmware}

Table~\ref{tab:zurich-specs} shows the firmware version of each Zurich Instruments
device used in this work.

\begin{table}[h]
  \center
  \begin{tabular}{lccc}
  \hline \hline
\textbf{Device}               & \textbf{Firmware}         & \textbf{HDL}         &  \\ \hline
    HDAWG Control             &    69121        &         69120           &  \\
    HDAWG Processing          &      69121       & 69080 &  \\
    PQSC                      &   69076            &   69076              &  \\
    SHFQC                     &  69120    & 69098   &  \\ \hline \hline
  \end{tabular}
\caption{Zurich FPGA internal controller software and HDL revision.}
\label{tab:zurich-specs}
\end{table}

\subsection{Cross-platform benchmark}\label{appendix:benchmark}

In this section we provide some more details on the experiments performed for the
performance benchmark presented in Sec.~\ref{sec:benchmark}.
A more detailed description of these routines is given by~\cite{PRXQuantum.2.040202,naghiloo2019introduction,48651}.
All these experiments were repeated for $4096$ shots.
For spectroscopies, a relaxation time of $5\,\mu$s was used,
while for the other experiments it was set at $300\,\mu$s.
Relaxation time is the waiting time between consecutive shots to let the qubit
relax back to the ground state before the next shot is started.

\begin{description}
    \item[Resonator spectroscopy] consists of a single-tone spectroscopy where a pulse is sent through the readout line and acquired through the feedback line. The frequency of the pulse is swept in a specific range, in our case probing $20$ or $100$ different frequencies. In the calibration of a 3D (2D) resonator, the amplitudes acquired present a positive (negative) peak at the resonance frequency of the resonator.
    \item[Qubit spectroscopy] consists of a two-tone spectroscopy where a first pulse is sent to the drive line and a measurement (a readout pulse and an acquisition) is performed right after. The frequency of the drive pulse is swept in a specific range. In the example used for the benchmark, $300$ frequencies were analyzed. As per the resonator spectroscopy, the amplitude acquired presents a peak for a specific frequency that, in this case, will be used as the drive pulse frequency.
    \item[Rabi amplitude] first a drive pulse, at the frequency identified with qubit spectroscopy, is sent through the drive line and a measurement is performed right after. The amplitude of the first pulse is swept in a range composed of, in this case, $75$ points. This experiment is used to calibrate the amplitude of the pi-pulse (Pauli-X gate) which rotates the qubits from the $\left | 0 \right \rangle$ state to $\left | 1 \right \rangle$.
    \item[Ramsey detuned] a first pulse is sent through the drive line. Then, after a delay, a new drive pulse is sent with a delay dependent phase and finally a measurement is performed. The delay between the two drive pulses, and therefore the phase, are swept. This experiment is used to fine tune the drive pulse frequency.
    \item[T1 experiment] the qubit is excited using a calibrated pi-pulse, then measured after a variable time. The characteristic decay shown by this experiment is used to measure the relaxation time T1 of the qubit.
    \item[T2 experiment] this experiment is almost identical to the Ramsey detuned experiment, but no additional phase is introduced in the second drive pulse. This enables to compute the characteristic dephasing time T2.
    \item[Single shot classification] The qubit is first just measured at the initial $\left | 0 \right \rangle$ state, and then excited and measured in the $\left | 1 \right \rangle$ state. The results are used to calibrate the classification between measured states.
    \item[Standard RB] First, a certain number (iterations) of circuits composed of Clifford gates is randomly generated. These circuits are executed and an average fidelity is computed. Then, new circuits are generated with increased depth and the procedure is repeated. The fidelity is supposed to decrease exponentially with the number of gates per circuit, leading to an estimation of the average error per gate.
\end{description}

\end{document}